\documentclass[conference]{IEEEtran}
\IEEEoverridecommandlockouts
\usepackage{cite}
\usepackage{amsmath,amssymb,amsfonts}
\usepackage{algorithm}
\usepackage{algpseudocode}
\usepackage{booktabs}
\usepackage{graphicx}
\usepackage{textcomp}
\usepackage{xcolor}
\usepackage{subfigure}
\usepackage{color, soul}
\usepackage{url,hyperref}
\def\BibTeX{{\rm B\kern-.05em{\sc i\kern-.025em b}\kern-.08em
    T\kern-.1667em\lower.7ex\hbox{E}\kern-.125emX}}
\begin{document}

\title{Accelerating Codec-based Speech Synthesis with Multi-Token Prediction and Speculative Decoding}
% MTCodec, MultiTokenCodec

\author{\begin{tabular}{c}
\textit{Tan Dat Nguyen$^{1}$, Ji-Hoon Kim$^{1}$, Jeongsoo Choi$^1$, Shukjae Choi$^2$, Jinseok Park$^2$, Younglo Lee$^2$, Joon Son Chung$^1$}\\
$^1$Korea Advanced Institute of Science and Technology, South Korea,
$^2$42dot Inc, South Korea \\
\{tandat.kaist, jh.kim, jeongsoo.choi, joonson\}@kaist.ac.kr \\
\{shukjae.choi, jinseok.park, younglo.lee\}@42dot.ai
\end{tabular}
}

% }

\maketitle

\begin{abstract}

The goal of this paper is to accelerate codec-based speech synthesis systems with minimum sacrifice to speech quality.
We propose an enhanced inference method that allows for flexible trade-offs between speed and quality during inference without requiring additional training.
Our core idea is to predict multiple tokens per inference step of the AR module using multiple prediction heads, resulting in a linear reduction in synthesis time as the number of heads increases.
Furthermore, we introduce a novel speculative decoding technique that utilises a Viterbi-based algorithm to select the optimal sequence of generated tokens at each decoding step. In our experiments, we demonstrate that the time required to predict each token is reduced by a factor of 4 to 5 compared to baseline models, with minimal quality trade-off or even improvement in terms of speech intelligibility. Audio samples are available at: \href{https://multpletokensprediction.github.io/multipletokensprediction.github.io/}{multpletokensprediction.github.io/multipletokensprediction.github.io/}.
\end{abstract}


\section{Introduction}
\label{sec:intro}
%%%%%%%%%%%%%%%%%%%%%%%%%%%%%%%%%%%%%%%%
% Overview Large language model and its application on speech synthesis$
%%%%%%%%%%%%%%%%%%%%%%%%%%%%%%%%%%%%%%%%%
Text-to-Speech (TTS)\cite{tan2021survey} has garnered significant attention within the research community. Amongst the various approaches, deep neural networks have shown considerable advancements in modeling natural speech\cite{oord2016wavenet,ryan2018tacotron,ren2019fastspeech,kim21vits,popov2021grad,wang23valle}.
Recent models based on this framework have shown the potential to generate fluent speech and replicate a speaker’s voice only using a few seconds of audio input~\cite{wang23valle,chen24valle2,du24vallt,han24valler,peng24voicecraft,song24ellav,Du24cosyvoice,borsos23soundstorm,zhang2023speechtokenizer,Yang23Uniaudio}.
Furthermore, LMs also possess outstanding ability on several speech task such as recognizing emotions, identifing speakers, enhancing speech, or even understanding the complex context from background sounds~\cite{Yang23Uniaudio,Nachmani24spectron, Zhang23speechgpt,Gong24listenthinkunderstanding}.
This ability is important to create an unified model that can naturally communicate with humans, reflecting emotion, accent, and style~\cite{Zhang23speechgpt}.

%%%%%%%%%%%%%%%%%%%%%%%%%%%%%%%%%%%%%%%%%%
% Overview of how this model work.
%%%%%%%%%%%%%%%%%%%%%%%%%%%%%%%%%%%%%%%%%%
The LM-based TTS model exhibits outstanding generative ability due to the capacity of the Transformer, as described by the scaling law~\cite{Kaplan20scalinglaw}, the advanced development of audio tokenizers~\cite{zhang2023speechtokenizer,Zeghidour22soundstream,Defossez23Encodec,Ju24naturalspeech}, and the avalability of large speech dataset~\cite{Nagrani17VoxCeleb,kang2023libriheavy,commonvoice:2020,Zen19libritts}.
This class of TTS models can be viewed as an extension of large language models, where the input speech is considered as a new language. The vocabulary of this new language is the set of discrete representations, also referred to as tokens, which are generated by encoder of audio codec. The TTS model then leverages its sequence generation ability to generate sequences of tokens that are transformed back to waveforms by the decodec module \cite{wang23valle,peng24voicecraft} or the vocoder module~\cite{Siuzdak24vocos,Lajszczak24basetts,kim21f_interspeech,Nguyen24FreGrad}. Since speech itself also has linguistic content in the form of a sequence, a promising approach is to harness the power of pretrained LLMs on text-only domains~\cite{Nachmani24spectron,Rubenstein23audiopalm,yang2024uniaudio15largelanguage}.

%%%%%%%%%%%%%%%%%%%%%%%%%%%%%%%%%%%%%%%%%%%
% Problem setup
%%%%%%%%%%%%%%%%%%%%%%%%%%%%%%%%%%%%%%%%%%%

Although AR inference brings significant improvements to codec-based speech synthesis, however it substantially reduces the generation speed of TTS systems. This slowdown is a notable drawback compared to other types of speech synthesis models. For example, non-AR models~\cite{ren2019fastspeech,kim21vits} generate an entire sentence in a single step, and models using progressive inference require only a few function evaluations (NFE) to produce high-quality speech output~\cite{popov2021grad,Huang22fastdiff,Ye23comospeech}. In contrast, models using AR inference with a language model (LM) architecture must generate sequences step by step, with the number of steps being equal to the sequence length. Consequently, as the sequence length increases, the time required for generation also increase. This problem is compounded by the fact that the computational complexity of Transformer-based models increases quadratically with sequence length, leading to an increase in the total floating-point operations per sequence.

\begin{figure*}
    \centering
    \subfigure[Training Process]{
        \includegraphics[width=0.88\columnwidth]{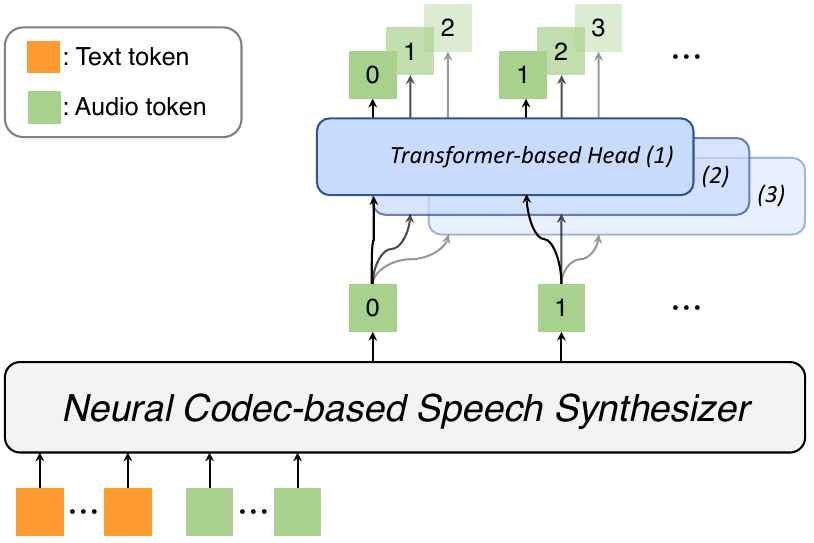}
        \label{fig:overall}}
    \subfigure[Inference Process]{
        \includegraphics[width=0.88\columnwidth]{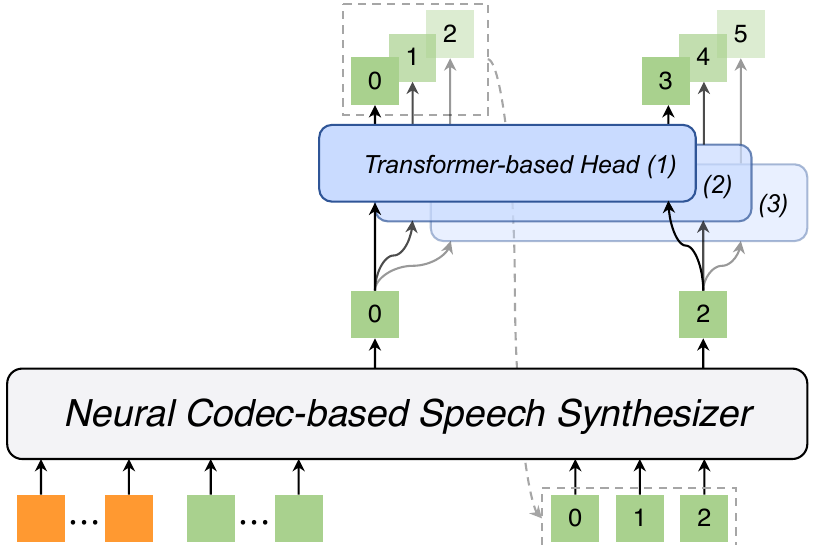}
        \label{fig:main-1}}
    \caption{AR module with $n = 3$ addition heads. Multiple heads are attached into current model that are simultaneously optimized to predict few future tokens given current state. Note that all the heads are trained once but can be flexibly ignored during inference for faster inference speed. During inference, few future tokens are predicted given current state per inference step. The generated tokens are concatenated into input sequence for the next inference step, respectively. Speculative decoding is also applied to enhance to quality of output tokens sequences.}
    \label{fig:main_architecture}
\end{figure*}

%%%%%%%%%%%%%%%%%%%%%%%%%%%%%%%%%%%%%%%
% Brief on method
%%%%%%%%%%%%%%%%%%%%%%%%%%%%%%%%%%%%%%%
In this paper, we introduce a multi-token prediction method that significantly reduces the NFE while maintaining comparable quality. Building on the observation that consecutive speech tokens are often similar, our architecture is trained to predict multiple future tokens simultaneously, rather than just one at a time. Additionally, we propose a Viterbi-like approach to capture the statistical relationships between the predicted tokens. A key advantage of our method is that it allows users to explicitly control the quality-speed trade-off by adjusting the number of future tokens predicted per function evaluation, without the need for re-training or fine-tuning the model. To our knowledge, this is the first instance of Viterbi-based speculative decoding in this context.

\section{Related Works}
\label{sec:related_works}
\textbf{Neural Audio Codec:} A codec model consists of an encoder, a decoder, and multiple codebooks that are quantized from the latent space of the encoder. These models are often built on the RVQ-GAN framework~\cite{Zeghidour22soundstream,Defossez23Encodec} which provide flexible bitrate for the task of audio compression. Recently, several approaches have been proposed for designing an optimal discrete representation space. Early works focused on improving compression and reconstruction quality~\cite{Defossez23Encodec,Zeghidour22soundstream,wang23valle}. As codec-based TTS systems have demonstrated strong performance, more recent efforts have focused on developing factorized and distinguishable features for each codebook~\cite{zhang2023speechtokenizer,Ju24naturalspeech,langman2024spectralcodecsspectrogrambasedaudio,ye2024codecdoesmatterexploring,liu2024semanticodec,ji2024wavtokenizer,Du24cosyvoice}. Moreover, current studies ~\cite{kim24clamtts,bai2024dmelspeechtokenizationsimple,langman2024spectralcodecsspectrogrambasedaudio,TakidaI24hqvae} introduce codecs for mel-spectrograms, significantly increasing token compression rates and leading to faster AR module inference. 
% Notably, our model not only adheres to these approaches, but also enhances them.

\textbf{Multi-tokens prediction for AR module:}
Multi-token prediction is not a new concept and can be related to speculative decoding~\cite{leviathan2023fast,chen2023accelerating,Cai24medusa}. These methods use a smaller draft model to generate an initial token sequence, which is then refined by the larger or original model for a coherent continuation. Recent works explore this approach in large language models (LLMs) like incorporating techniques to reduce hardware memory requirements~\cite{Gloeckle24betterandfaster} or using tree search algorithms to enhance the quality of the output sequence~\cite{Cai24medusa}. To our knowledge, the most related work to ours is VALL-E 2~\cite{chen24valle2}, but the multi-token prediction methods used in these two approaches are fundamentally different. Additionally, our work not only offers explicit, training-free control over the quality-speed trade-off but also improves the relationship between the predicted tokens through a Viterbi-like algorithm.

\section{Method}
\label{sec:method}
\subsection{Problem modeling}

The popular codec-based speech synthesis model architecture comprises two main modules: an auto-regressive (AR) module and a non-auto-regressive (NAR) module, following the VALL-E framework~\cite{wang23valle}. Specifically, given a training dataset with pairs ${\boldsymbol{s}, \boldsymbol{p}}$, where $\boldsymbol{s}$ represents the speech signal and $\boldsymbol{p}=\{p_1, p_2, \ldots, p_T\}$ is the corresponding phoneme sequence of length $T$, the codec model compresses the speech signal into discrete tokens $\boldsymbol{A}$ using eight quantizers: $Codec(\boldsymbol{s}) = \boldsymbol{A}^{8\times L} = \{\boldsymbol{a}^1, \boldsymbol{a}^2, \ldots, \boldsymbol{a}^8\}$, where each $\boldsymbol{a}^i = \{a^i_1, a^i_2, \ldots, a^i_L\}$. In models using the VALL-E architecture, the AR module is responsible for predicting $\boldsymbol{a}_1$, while the NAR module regresses the entire sequence $a^i\big|_{i=2}^8$.

\begin{figure*}[t]
  \centering
  \begin{minipage}{0.63\textwidth}
    \centering
    \includegraphics[width=\textwidth]{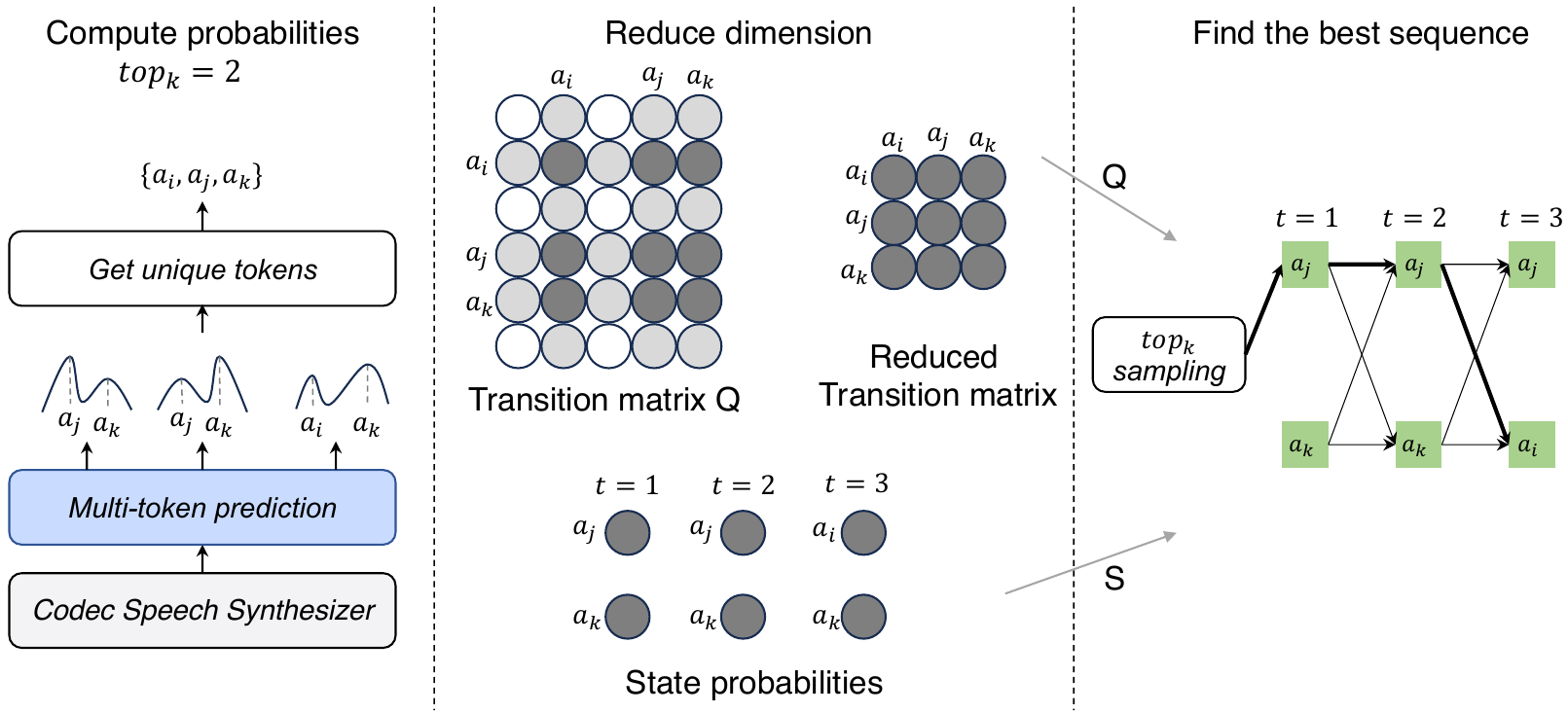}
    \caption{Viterbi-based Speculative Decoding is illustrated as follows: (1) Multiple prediction heads generate several distributions per timestep simultaneously. (2) To optimize memory and computational efficiency, the dimensions of the transition matrix and state probabilities are reduced by selecting only the necessary rows and columns. (3) The best sequence is determined using Speculative Decoding, as described in Algorithm \ref{algo:speculative_decoding}. The transition matrix computation for LibriTTS is completed in just 3 minutes. Additionally, $top_k$ sampling is employed to preserve diversity.}
    \label{fig:viterbi}
  \end{minipage}%s
  \hfill
  \begin{minipage}{0.35\textwidth}
    \centering
    \input{algos/viterbi}
  \end{minipage}
\end{figure*}

% \begin{figure*}
%     \centering
%     \includegraphics[width=0.9\linewidth]{images/Viterbi.pdf}
%     \caption{Speculative decoding by Viterbi in $top_k = 2$ predictions}
%     \label{fig:enter-label}
% \end{figure*}

Specifically, for the discrete token sequence $\boldsymbol{a}^1$, an auto-regressive Transformer Decoder-only $\boldsymbol{\theta}_{AR}$ is trained to predict the next tokens based on a text prompt $\boldsymbol{p}$, and an acoustic prompt condition $\boldsymbol{\Tilde{a}}^1$ extracted from the reference audio, and all previously predicted tokens.

\begin{equation}
\label{eq:ar}
p(\boldsymbol a^1 \vert \mathbf{p}, \boldsymbol{\Tilde{a}}^1; \boldsymbol{\theta}_{AR}) = \prod_{t=0}^{L} p(\boldsymbol{a}^1_{t+1} \vert \boldsymbol{a}^1_{<t}, \tilde{\mathbf{a}}^1, \mathbf{p}; \boldsymbol{\theta}_{AR})
\end{equation}

Since all data are discrete, we concatenate them all into a single sequence without specific tokens to distinguish them.
For tokens of the remain layers $a^i\big|_{i=2}^8$, we train a NAR LM $\boldsymbol{\theta}_{NAR}$ to iteratively predict each token sequence. Each prediction process is conditioned on phoneme sequence $\boldsymbol{p}$ and generated token sequence of previous layers $\boldsymbol{a}^{<j}$.

\begin{equation}
\label{eq:ar}
p(\boldsymbol a^{2:8} \vert \mathbf{p}, \boldsymbol{\Tilde{a}}; \boldsymbol{\theta}_{NAR}) = \prod_{j=2}^{8} p(\boldsymbol{a}^{j} \vert \boldsymbol{a}^{<j}, \tilde{\mathbf{a}}, \mathbf{p}; \boldsymbol{\theta}_{NAR})
\end{equation}

\subsection{Multiple Tokens Prediction (MTP)}
We take inspiration from \cite{Cai24medusa,Gloeckle24betterandfaster}, which utilizes parallel decoding to accelerate the decoding process of LLM. We divide the AR module into two parts: first, encode all previous tokens in to latent space $p(\boldsymbol{z}_{1:t}\vert \boldsymbol{a}^1_{1:t}, \tilde{\boldsymbol{a}}, \boldsymbol{p}; \boldsymbol{\theta}_{AR})$, then, using this latent $\boldsymbol{z}_{1:t}$ to predict a few $n$ consecutive future tokens $p(\boldsymbol a^1_{t+1:t+n}\vert \boldsymbol z_{1:t})$ by  $n$ 
 distinguished heads. This idea leads to a new negative log-likelihood objective function of the whole AR module with multiple heads:
\begin{align}
\mathcal{L}  &= -\sum_{t} \sum_{i=1}^{n} \log P(a^1_{t+i} \vert \boldsymbol{z}_{1:t}) P(\boldsymbol{z}_{1:t}\vert\boldsymbol{a}^1_{1:t}, \tilde{\boldsymbol{a}}, \boldsymbol{p}) \\
\label{eq:multihead_loss}
&\ge -\sum_{t} \log P(\boldsymbol{a}^1_{t+1:t+n}\vert z_{1:t})  P(\boldsymbol{z}_{1:t}\vert\boldsymbol{a}^1_{1:t}, \tilde{\boldsymbol{a}}, \boldsymbol{p}) \\
&= -\sum_{t} \log P(\boldsymbol{a}^1_{t+1:t+n}\vert\boldsymbol{a}^1_{1:t}, \tilde{\boldsymbol{a}}, \boldsymbol{p})
\end{align}

The equality in \ref{eq:multihead_loss} holds when $n$ consecutive tokens are independent given $\boldsymbol{z}_{1}$. This highlights the need for a tokenizer that supports such inference. To the best of our knowledge, SpeechTokenizer~\cite{zhang2023speechtokenizer} one is the most suitable candidate, as its first-layer tokens encode only linguistic information, effectively serving as pseudo labels for phonemes. This makes future token prediction easier compared to Encodec~\cite{Defossez23Encodec}, where linguistic and acoustic information are entangled at every quantizer layer. Furthermore, by applying $n$ heads in parallel during inference, we reduce the time complexity from $\mathcal{O}(L)$ to $\mathcal{O}(\frac{L}{n} + \alpha)$, where $\alpha$ represents the additional computational overhead introduced by the extra heads.

\subsection{Vitebi-based Speculative Decoding}

Predicting multiple tokens $\boldsymbol{a}^1_{t+1:t+n}$ at once, based on the context from $\boldsymbol{a}^1_{\le t}$, introduces a context mismatch among the predicted tokens. Specifically, except for $\boldsymbol{a}^1_{t+1}$, all tokens $\boldsymbol{a}^1_{t+2:t+n}$ are predicted without full awareness of the preceding tokens, which can lead to incorrect predictions that accumulate over time and compromise the entire sequence. In this work, we propose a Viterbi-based speculative decoding technique to bridge the contextual gap between successive tokens predicted simultaneously. The key objective of this algorithm is to select the best possible sequence given $top_k$ tokens of the $n$ heads.

% SETUP MARKOV CHAIN
Let $V$ represent the set of tokens from the first layer of the RVQ model. We assume each token depends only on its previous token, allowing us to model the token transitions using a Markov chain. Our goal is to model the likelihood of token $a_i$ occurring immediately before token $a_j$, represented by the probability $\boldsymbol{Q}(i,j)$, where $\boldsymbol{Q} \in \mathbb{R}^{\vert V\vert \times \vert V\vert}$ is the transition matrix derived from the dataset. Additionally, we obtain $S_t(a_i)$ from the $t$ heads of the multi-token prediction (MTP), representing the probability of $a_i$ at time $t$. Therefore, the probability of transitioning from $a_i$ at time $t-1$ to $a_j$ at time $t$ is $Q(a_i, a_j) \cdot S_t(a_j)$.

To determine the optimal sequence of tokens $A^*$, we apply Viterbi-based speculative decoding, as shown in Fig. \ref{fig:viterbi} and Algorithm \ref{algo:speculative_decoding}. However, implementing this algorithm has a computational complexity of $\mathcal{O}(n * V^2)$, leading to high computational costs. To mitigate this, we reduce the dimensions of both the transition matrix and the state matrix by selecting only the relevant elements during each inference step. This reduces the size of the transition matrix from $\mathbb{R}^{\vert V\vert \times \vert V\vert}$ to $\mathbb{R}^{m \times m}$, where $m \leq k \cdot n$ is the total number of unique tokens from the top $k$ highest-probability predictions across the $n$ heads of the MTP. In practice, $m$ is much smaller than $V$, resulting in a new computational complexity of $\mathcal{O}(n * m^2)$, which significantly reduces the overhead computation time for each MTP inference.

\begin{table*}[t]
\centering
\caption{Experimental results on LibriTTS dataset. TPT denotes time per token (ms/token) and Speedup refers to the extent of speed improvement compared to the baseline.
MOS is presented with 95\% confidence interval. $\uparrow$: higher is better, $\downarrow$: lower is better.}
\begin{tabular}{l|cc|cc|cccc}
\toprule
Model & TPT$\downarrow$ & Speedup$\uparrow$ & MOS$\uparrow$ & SMOS$\uparrow$ & UTMOS$\uparrow$ & SIM$\downarrow$  & WER(\%)$\downarrow$  & CER(\%)$\downarrow$    \\ \midrule
Ground truth & -- & -- &4.77$\pm$0.09 &3.53$\pm$0.29 &4.14 &-- &~2.9 &0.8 \\ \midrule
VALL-E~\cite{wang23valle} &20.5 &-- &{\bf 2.89$\pm$0.17} & {\bf 3.20 $\pm$0.20} &{\bf 3.56} &{\bf 79.6} &12.8 &{\bf 8.1} \\
~~~+ {\bf Ours} &{\bf 4.5} &$\times$4.56 &2.79$\pm$0.18  &2.79$\pm$0.20 &3.53 &79.2 &{\bf 12.0} &8.3 \\ \midrule
USLM~\cite{zhang2023speechtokenizer} &23.5 & -- &3.72$\pm$0.15 &3.00$\pm$0.20 &{\bf 3.82} &{\bf 79.3} &11.5 &7.7\\
~~~+ {\bf Ours} &{\bf 4.4} &$\times$5.34 &{\bf 4.02$\pm$0.12} &{\bf 3.25$\pm$0.19} &{\bf 3.82} &78.8 &~{\bf 8.7} &{\bf 5.2}  \\
\bottomrule
\end{tabular}
\label{tab:main}
\end{table*}

\begin{figure*}[h]
  \centering
  \begin{minipage}{0.24\textwidth}
    \centering
    \includegraphics[width=\textwidth]{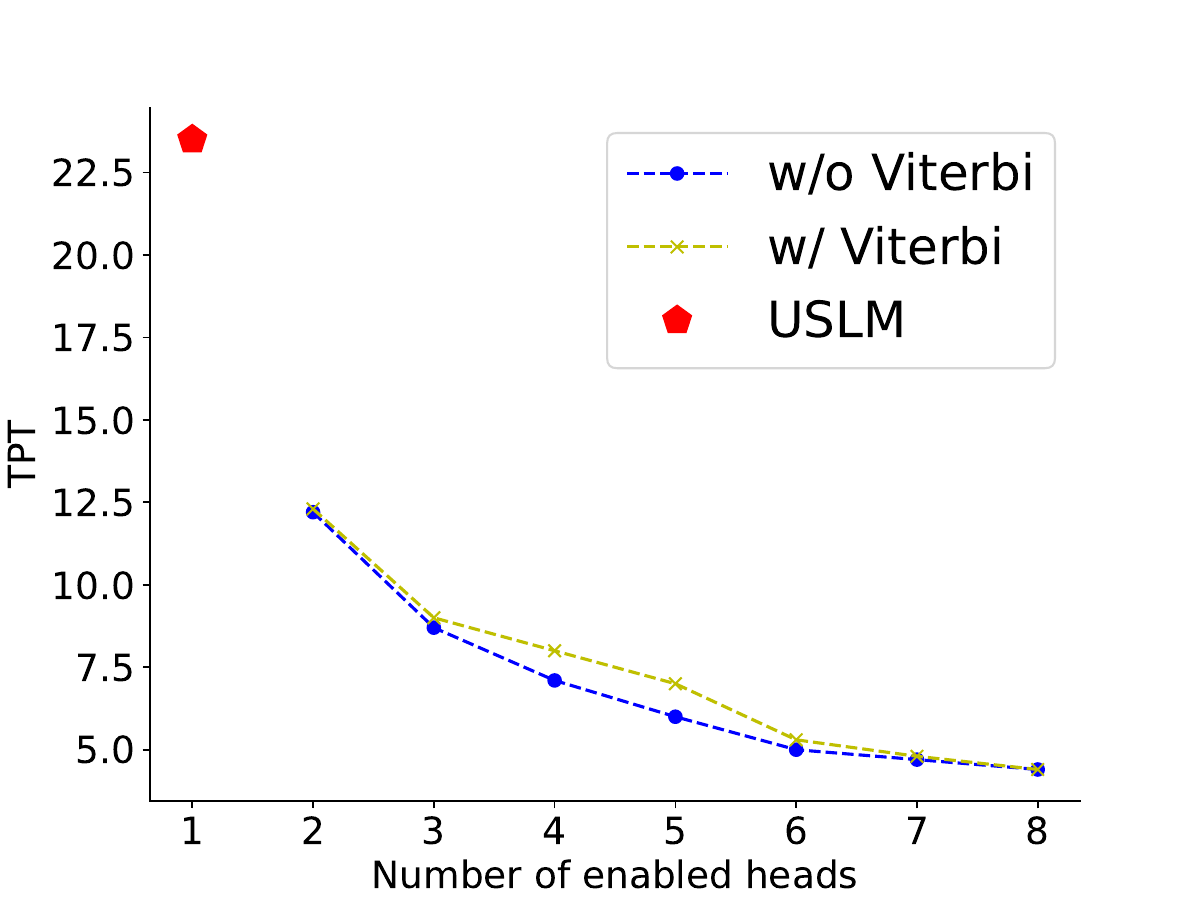}
    \caption{TPT$\downarrow$ over num. of heads}
    \label{fig:tpt}
  \end{minipage}%
  \hfill
  \begin{minipage}{0.24\textwidth}
    \centering
    \includegraphics[width=\textwidth]{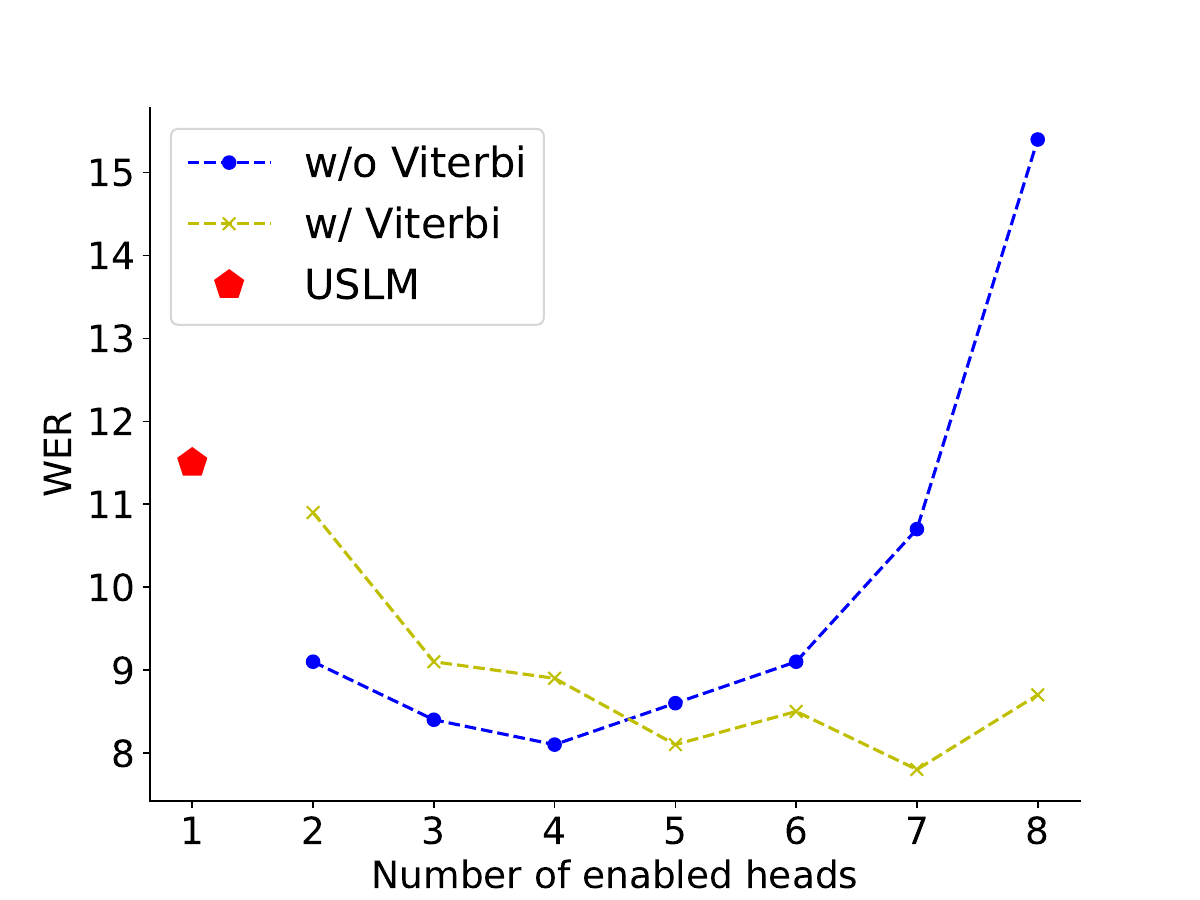}
    \caption{WER$\downarrow$ over num. of heads}
    \label{fig:wer}
  \end{minipage}
  \hfill
  \begin{minipage}{0.24\textwidth}
    \centering
    \includegraphics[width=\textwidth]{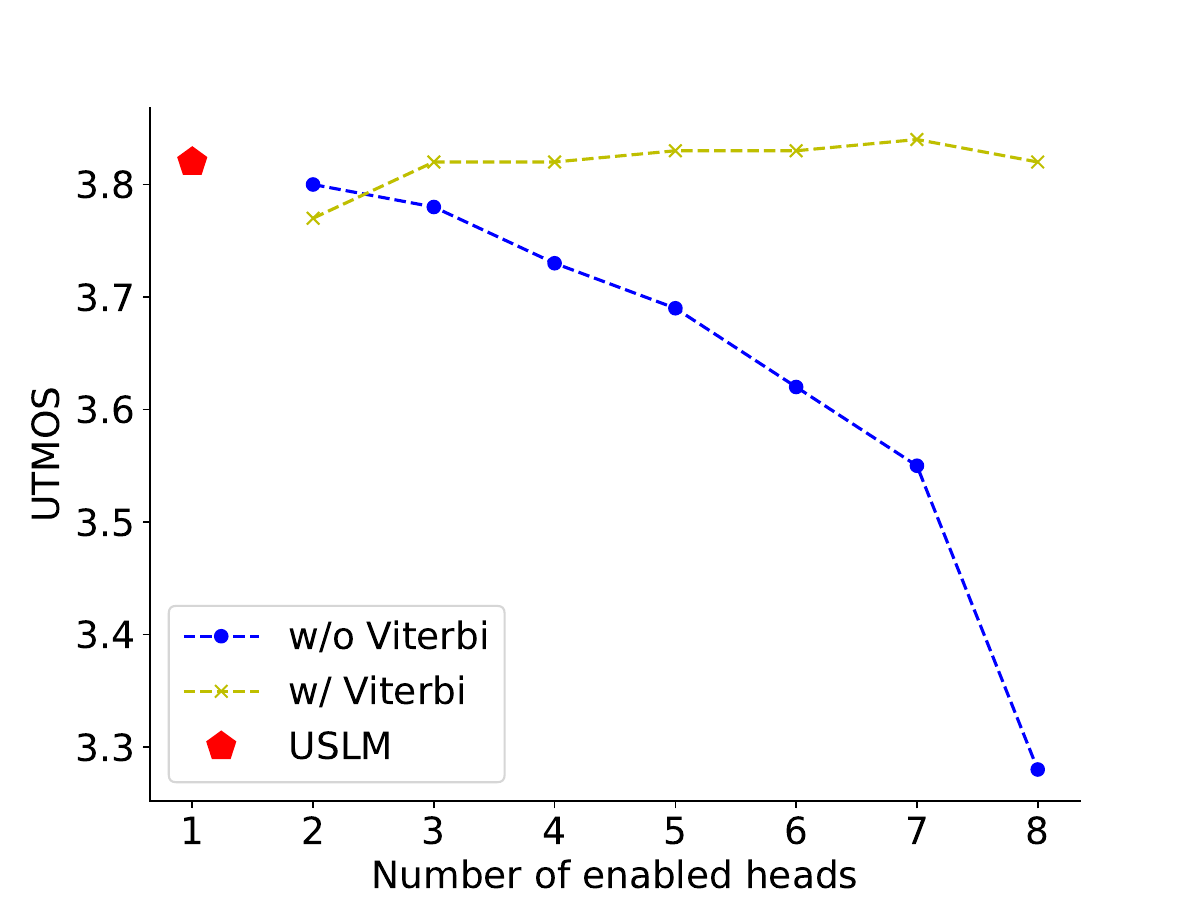}
    \caption{UTMOS$\uparrow$ over num. of heads}
    \label{fig:utmos}
  \end{minipage}\hfill
  \begin{minipage}{0.24\textwidth}
    \centering
    \includegraphics[width=\textwidth]{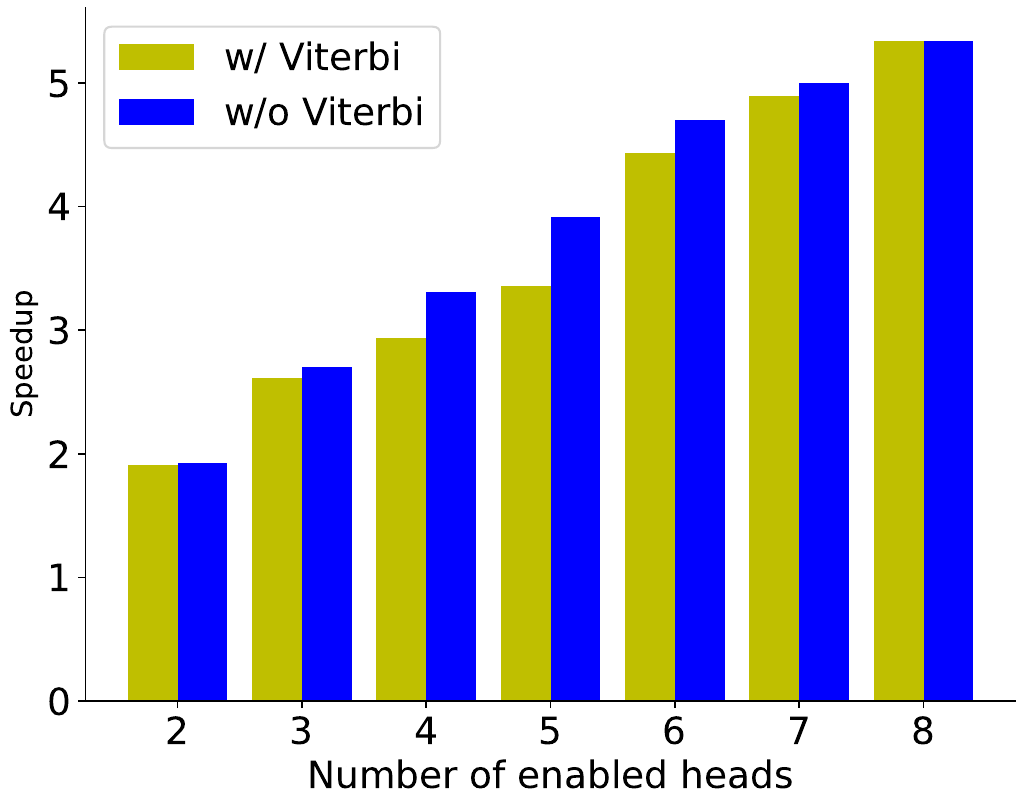}
    \caption{Speedup ratios of MTP }
    \label{fig:speedup}
  \end{minipage}
  % \caption{Ablation studies on how Viterbi affect the content and quality consistency}
\end{figure*}

% \begin{table}[h]
% \centering
% \caption{Experimental results on LibriTTS dataset. TPT denotes time per token (ms/token) and Speedup refers to the extent of speed improvement compared to the baseline.
% MOS is presented with 95\% confidence interval. $\uparrow$: higher is better, $\downarrow$: lower is better.}
% \begin{tabular}{r|c|c|c}
% % \multicolumn{4}{c}{\textbf{Performance Metrics Summary}}
% \toprule
% $top_k$ & TPT$\downarrow$ & UTMOS$\uparrow$ &  WER(\%)$\downarrow$     \\ \midrule
% 3 & 4.4 & 3.82 & 8.6 \\
% 5 & 4.4 & 3.79 & 9.3  \\
% 7&4.4 & 3.76 &9.7  \\
% 9&4.4 &3.74 &10.2 \\
% 15&4.5 & 3.67 &11.1  \\
% 25&4.5 & 3.58 &12.2  \\
% \bottomrule
% \end{tabular}
% \label{tab:main}
% \end{table}

\begin{table}[h]
\centering
\caption{Ablation studies on the effect of $top_k$ in speed and quality of multiple token prediction}
\label{tab:ablation_topk}
\begin{tabular}{l|ccc|ccc}
\toprule
 &\multicolumn{3}{c}{$n = 8$} & \multicolumn{3}{c}{$ n = 4$} \\ \midrule
$top_k$   & TPT$\downarrow$ & UTMOS$\uparrow$ &  WER$\downarrow$ & TPT$\downarrow$ & UTMOS$\uparrow$ &  WER$\downarrow$       \\ \midrule
3 & 4.4 & 3.82 & 8.6 & 8.0 & 3.82 & 8.1 \\
5 & 4.4 & 3.79 & 9.3 & 8.0 & 3.78 & 8.7 \\
7&4.4 & 3.76 &9.7 & 8.0 &  3.72 & 9.9  \\
9&4.4 &3.74 &10.2 & 8.0 & 3.65 & 10.9 \\
15&4.5 & 3.67 &11.1 & 8.0 & 3.49 & 12.2\\
25&4.5 & 3.58 &12.2 & 8.1 & 3.31 & 13.9 \\
\bottomrule
\end{tabular}
\end{table}
\section{Experiments}
\label{sec:exp}

\subsection{Dataset}

We conduct experiments on the LibriTTS~\cite{Zen19libritts} corpus, using the all train subsets for training, and 'test-clean' for evaluation, following \cite{Du24cosyvoice}. We also use SpeechTokenizer to extract audio tokens all experiments. For fair comparison, all models are trained from scratch. Both AR and NAR modules are optimized separately for 20 epochs on a single A6000 GPU, with a learning rate of 0.05 and 200 warmup steps.

\subsection{Comparision with baselines}

We integrate our method into state-of-the-art models, VALL-E~\cite{wang23valle} and USLM~\cite{zhang2023speechtokenizer}, and evaluate their zero-shot performance using various metrics. For content consistency, we compute Word Error Rate (WER) and Character Error Rate (CER) with a Conformer-Transducer speech-to-text system\footnote{nvidia/teams/nemo/models/stt\_en\_conformer\_transducer\_large}. Speaker similarity is assessed via cosine similarity of $x$-vector embeddings between prompts and generated utterances. We also conduct SMOS and MOS surveys for speaker similarity and audio quality (5-point scale), and use UTMOS~\cite{Saeki22UTMOS} for MOS estimation. Generation efficiency is measured using time-per-token (TPT). All metrics, except the MOS and SMOS surveys (50 random samples per model), are computed on the test-clean set. Results are shown in Table \ref{tab:main}.

% QUality comparision
Compared to the baselines, multiple token prediction with 8 prediction heads and speculative decoding accelerates the inference process by 4 to 5 times while maintaining comparable speech quality and speaker similarity. Notably, content consistency improves, with WER decreasing in both USLM and VALL-E, and CER improving in USLM. This demonstrates the potential of multiple token predictions to enhance content accuracy while speeding up the AR process. Specifically, USLM with speculative decoding achieved the best WER of $8.7\%$ and was the fastest among all models. The difference in inference time per token between VALL-E and USLM arises from the variance in the length of sequences generated by each model, which lead to variance of computing complexity. However, the speedup ratios are both theoretically and experimentally validated, ensuring that the performance improvements are consistent across different conditions.

We further assess the effectiveness of Viterbi-based speculative decoding by comparing quality and content consistency across different numbers of prediction heads. Figure \ref{fig:tpt} shows that increasing the number of heads accelerates the AR process, with minimal impact on speed due to the reduced transition matrix dimensions, which shrink the search space for the Viterbi algorithm. The slight overhead from speculative decoding is highlighted in Fig.~\ref{fig:speedup}, comparing scenarios with Viterbi enabled (w/ Viterbi) or disabled (w/o Viterbi). To examine the trade-off between speed and quality, we evaluate WER, CER, and UTMOS for prediction heads ranging from 2 to 8. Without Viterbi-based decoding, content consistency remains acceptable (WER) with a few heads, but UTMOS drops rapidly. With more than 4 heads, both speech quality and content accuracy degrade significantly because of accuracy reduction in far future heads. However, speculative decoding maintains consistent synthesis performance across different numbers of heads, as shown in Figs. \ref{fig:wer} and \ref{fig:utmos}. Notably, higher $top_k$ expands the Viterbi algorithm's search space with trivial additional time. According to Table \ref{tab:ablation_topk}, although higher $top_k$ increases diversity in generated speech, it tends to lower overall quality in terms of WER and UTMOS.

\section{Conclusion}
\label{sec:conclusion}

In this work, we propose an inference method that combines multi-token prediction with Viterbi-based speculative decoding, significantly accelerating the AR module during inference. Our approach not only achieves faster inference but also maintains, or even surpasses, baseline performance in terms of generation quality. Additionally, the model is trained only once, yet offers explicit control over the trade-off between quality and speed. However, exploring the compatibility of more tokenizers with Viterbi-based speculative decoding is a promising direction for future work.
\clearpage\newpage
\bibliographystyle{IEEEtran}
\bibliography{main}

\end{document}